\RequirePackage{fix-cm}
\documentclass[twocolumn]{svjour3}          % twocolumn

\smartqed  % flush right qed marks, e.g. at end of proof

\usepackage{graphicx}
\usepackage{dcolumn}
\usepackage{bm}
\usepackage{hyperref}
\usepackage{subfigure}
\hypersetup{colorlinks=true, citecolor=blue, urlcolor=blue, linkcolor=blue}
\usepackage{amssymb}
\usepackage{color}
\usepackage{amsmath}
\usepackage{mathrsfs}
\usepackage{times}
\usepackage{subeqnarray}
\usepackage{cases}
\usepackage{epstopdf}
\usepackage[numbers,sort&compress]{natbib} 
\usepackage{marvosym}
\usepackage{ifsym}

\makeatletter
\def\paragraph{\@startsection{paragraph}{4}%
  \z@\z@{-\fontdimen2\font}%
  {\normalfont\bfseries}}
\makeatother
\begin{document}

\title{Oscillatory cooperation prevalence emerges from misperception}
\author{Jing Zhang \and Zhao Li \and Jiqiang Zhang \and Lin Ma \and Guozhong Zheng \and Li Chen}

\institute{J. Zhang \and Z. Li \and L. Ma \and G. Zheng \and L. Chen(\Letter) \at
              School of Physics and Information Technology, Shaanxi Normal University, Xi'an 710062, P. R. China \\
              \email{chenl@snnu.edu.cn}          
           \and
           J. Zhang \at
              School of Physics and Electronic-Electrical Engineering, Ningxia University, Yinchuan 750021, P. R. China
}

\date{Received: date / Accepted: date}

\titlerunning{Oscillatory cooperation prevalence emerges from misperception}
%\authorrunning{J. Zhang et al.}

\maketitle

\begin{abstract}
Oscillatory behaviors are ubiquitous in nature and the human society. However, most previous works fail to reproduce them in the two-strategy game-theoretical models.  Here we show that oscillatory behaviors naturally emerge if incomplete information is incorporated into the cooperation evolution of a non-Markov model. Specifically, we consider a population playing prisoner's dilemma game, where each individual can only probabilistically get access to their neighbors' payoff information and store them within their memory with a given length. They make their decisions based upon these memories. Interestingly, we find that the level of cooperation generally cannot stabilize but render quasi-periodic oscillation, and this observation is strengthened for a longer memory and a smaller information acquisition probability. The mechanism uncovered shows that there are misperceived payoffs about the player's neighborhood, facilitating the growth of cooperators and defectors at different stages that leads to oscillatory behaviors as a result. Our findings are robust to the underlying structure of the population. 
Given the omnipresence of incomplete information, our findings may provide a plausible explanation for the phenomenon of oscillatory behaviors in the real world.
\keywords{ Cooperation \and Evolutionary game theory \and  Prisoner's dilemma \and Oscillation}
\end{abstract}

\section{Introduction}
\label{sec:introduction}
Human civilization is the result of large-scale cooperation in every aspect of socioeconomic activities, ranging from raising offsprings to the division of labor ~\cite{axelrod1981evolution,smith1997major}. Altruistic cooperation behaviors are also widely observed in different ecosystems in the wild~\cite{clutton2009cooperation}. However, according to Darwin's \emph{The Origin of Species}, ``survival of the fittest" implies the innate selfish nature of individuals. This means that cooperative behaviors are actually counterintuitive, because cooperators helping others is at a certain cost, thus put themselves in an inferior competitive position to those free-riders~\cite{rider1984evolution,smith1982evolution,gintis2000game,nowak2006evolutionary}. As a result, deciphering the mechanism of cooperation becomes a fundamental challenge, which has attracted many researchers' attention from different fields and now is a highly interdisciplinary field~\cite{gintis2000game}.

To proceed, the evolutionary game theory \cite{nowak2004evolutionary} has been introduced and is proved to be a useful theoretical framework. By analysing stylized social dilemmas (e.g. the prisoner’s dilemma, the snowdrift game, and the public goods game), several mechanisms have been proposed \cite{nowak2006evolutionary}, such as direct~\cite{trivers1971evolution} and indirect reciprocity~\cite{zahavi1999handicap}, kin~\cite{hamilton1964genetical} and group selection~\cite{Keller0Levels,Queller1964Group}, reward and punishment~\cite{nowak2006five}, social diversity~\cite{sigmund2001reward} and social hierarchy~\cite{liang2021social},  spatial or network reciprocity \cite{nowak1992evolutionary}. Particularly, theoretically considering the fact that human populations are structured with fixed neighbors can support cooperation. The argument behind is that cooperator clusters are more likely to form in the structured population, which potentially are able to defeat the invasion of defectors, compared to the well-mixed scenario. However, human behavioral experiments do not support this theoretic prediction in general~\cite{traulsen2010human}. This unsatisfactory situation suggests that the evolution of cooperation in realistic population could be far more complex than what current game-theoretic models captured, and the experiment-driven modeling approach is needed. It's worthy noting that the dynamical reciprocity is recently also proposed as a counterpart mechanism~\cite{liang2022dynamical}, showing that the game-game interactions could also improve the cooperation prevalence, though remains to be validated in the experiments.

While most of previous works aim to understand the emergence of cooperation where the prevalence finally settles down, there are quite some systems where the oscillatory behaviors are often seen. The most prominent example is the population oscillation of different species in many ecological systems, which is vital to preserving biodiversity~\cite{kerr2002local,reichenbach2007mobility}. 
An early work~\cite{nowak1993chaos} shows that in a population consisting of some competing strategies, strategic mutation can lead to persistent periodic or even chaotic oscillations in the frequencies of the strategies and the level of cooperation. Although the model is formulated within the iterated prisoner's dilemma framework, the strategies are defined by a quadruple of parameters, where there are as much as 16 species are engaged in the evolution. 
Later, a mainstream explanation for oscillation is via the cyclic dominance~\cite{sinervo1996rock,szabo2002phase,kerr2002local,reichenbach2007mobility,imhof2005evolutionary,szolnoki2014cyclic}. A popular example is the Rock-Paper-Scissors game~\cite{zhou2016rock}, where the fractions of three strategies naturally oscillate given the population is spatially structured~\cite{szolnoki2010dynamically} or they are capable of mobility~\cite{szabo2002phase,kerr2002local,reichenbach2007mobility}. However, this sort of models require three or more species and thus cannot explain how the oscillatory behaviors emerge in the 2-specie populations, such as the classic 10-year population cycle of snowshoe hares and Canada lynx  in the boreal forests of North America{~\cite{elton1942ten,moran1953statistical,krebs2001drives}}. 
Recently, Yang et al.~\cite{yang2022oscillation} observed a tide-like burst in a probabilisitic migration model of prisoner's dilemma, where migration is driven by conformity and self-centered inequity aversion norms. 
It's also worth noting that recently Zhang et al.~\cite{zhang2020oscillatory} introduced the reinforcement learning framework, where periodic burst-like oscillation is also seen in pairwise games.

Here we provide a different explanation for oscillations in the pairwise game within the classic framework, where we incorporate two commonly seen ingredients: the non-Markov effect and incomplete information. While the former considers the historical impact, the latter is due to the limited information acquisition capability, because their neighbors' payoff information is not always accessible.
In fact, the argument behind the non-markov effect is well-acknowledged that the decision-making for individuals is based on the knowledge of the past records rather than a single round~\cite{szabo2007evolutionary}. Plenty of theoretical works show that memory can considerably improve the cooperation level~\cite{wang2006memory,qin2008effect,yong2010payoff,jiang2009reducing}, and several related factors such as the length of memory, the temptation of defection, and the structure of networks have been studied~\cite{luo2016cooperation,yong2010payoff,yang2012effects,stewart2016small,liu2020link,horvath2012limited,shu2019memory,ye2017memory,wang2020learning}.  
The impact of incomplete information scenario, where players do not possess full information about their opponents, has also been investigated separately in previous studies.
Ref.~\cite{vuolevi2012boundaries} shows experimentally that incompleteness of information about the opponent's past behaviors undermines both expectations about another person's cooperation as well as one's own cooperation. Similar observations are made that selfish behaviors are the best way of protecting oneself against non-cooperative behaviors in the presence of incomplete information~\cite{tazdait2008mutual}.
However, in a finitely repeated PD, incomplete information about their partners' preference is found to trigger the possibility of coopepration~\cite{dijkstra2017explaining}.

In this work, we find that when the two ingredients are incorporated simultaneously, oscillatory behaviors emerge, where the cooperation prevalence rises and falls in a quasi-periodic way. Specifically, we consider a structured population of individuals where they can only probabilistically get access to their neighbors' payoffs and store this information in their memory; by comparing the perceived payoffs to their own values they decide whether to imitate their neighbours' cooperation propensity or not. We find oscillation is likely to occur when the memory is long and the payoff information is hard to obtain. Further analysis shows that the oscillation is an intrinsic property of the system, the presence of incomplete information leads to a misperception of neighbors' payoff that causes cooperation explosion and decline. The longer memory and stronger incomplete information are, the stronger perception errors are expected.

This paper is organized as follows. We introduce our model in Sec. \ref{sec:model}. Numerical simulation results on the $2d$ square lattice are shown in Sec. \ref{sec:Numerical results}. The mechanism is analysed in Sec. \ref{sec:Mechanism analysis}. Finally, we conclude and discuss our work in Sec. \ref{sec:conclusion}.

\begin{figure*}[tbp]
\centering
\includegraphics[width=0.45\linewidth]{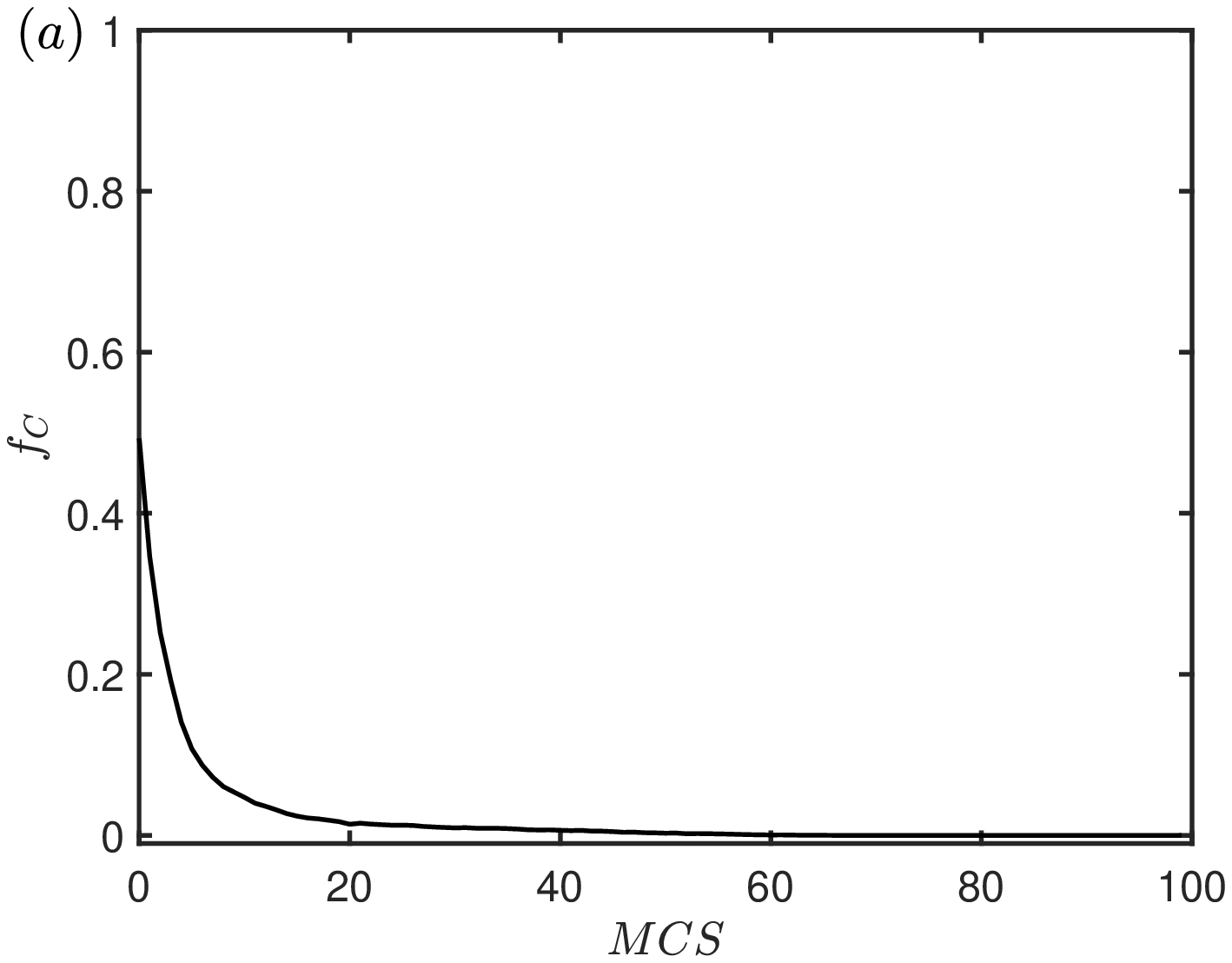}
\includegraphics[width=0.45\linewidth]{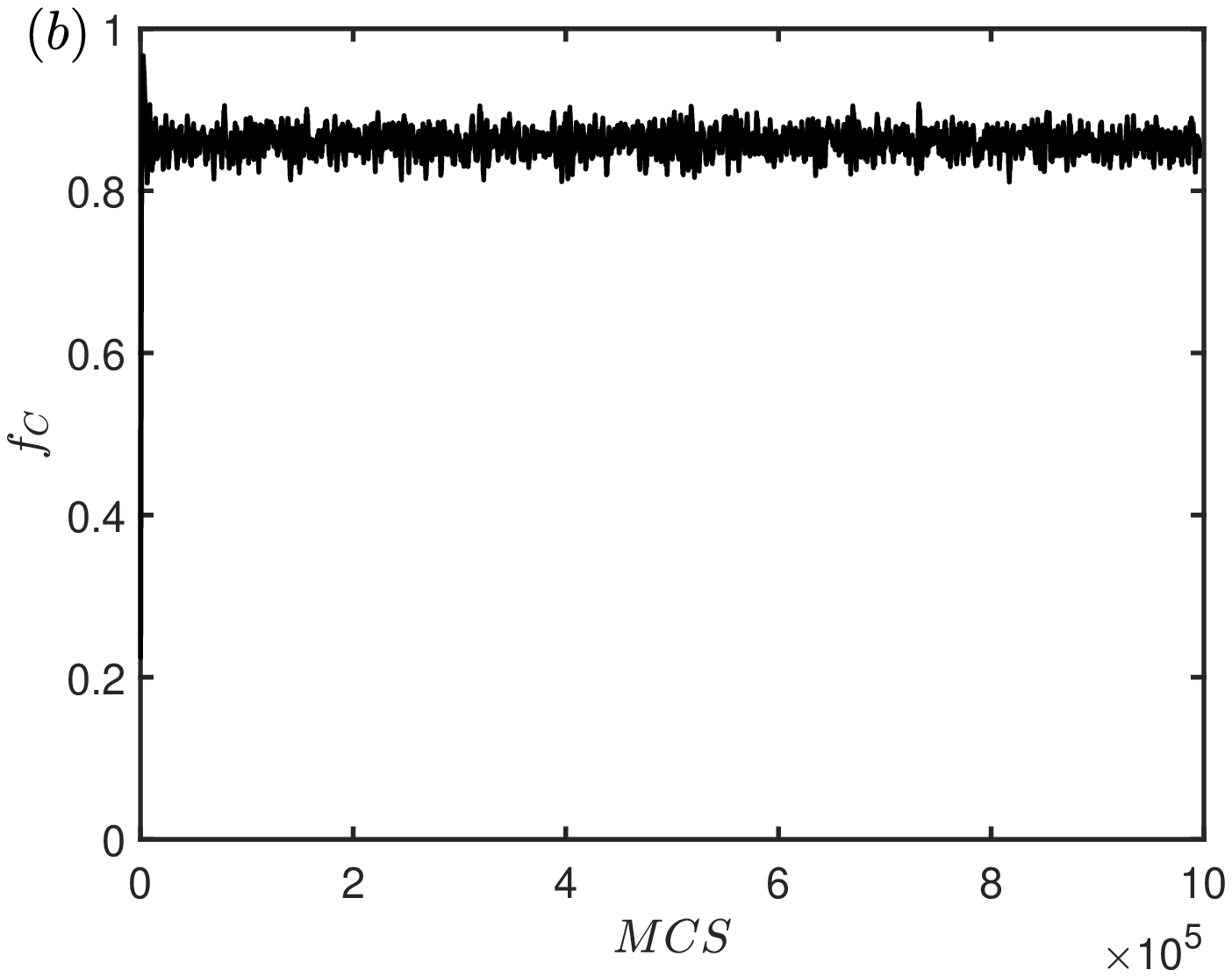}
\includegraphics[width=0.45\linewidth]{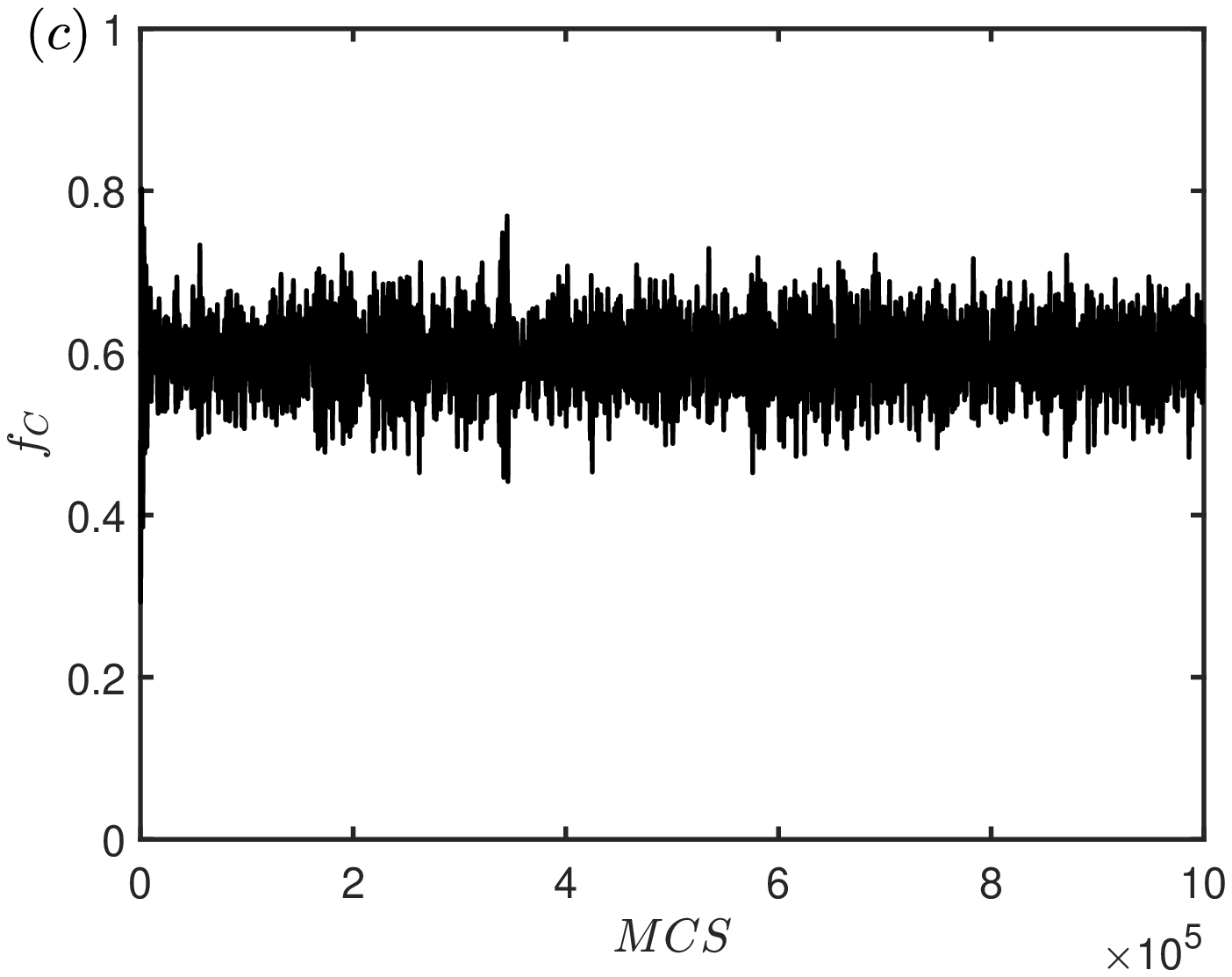}
\includegraphics[width=0.45\linewidth]{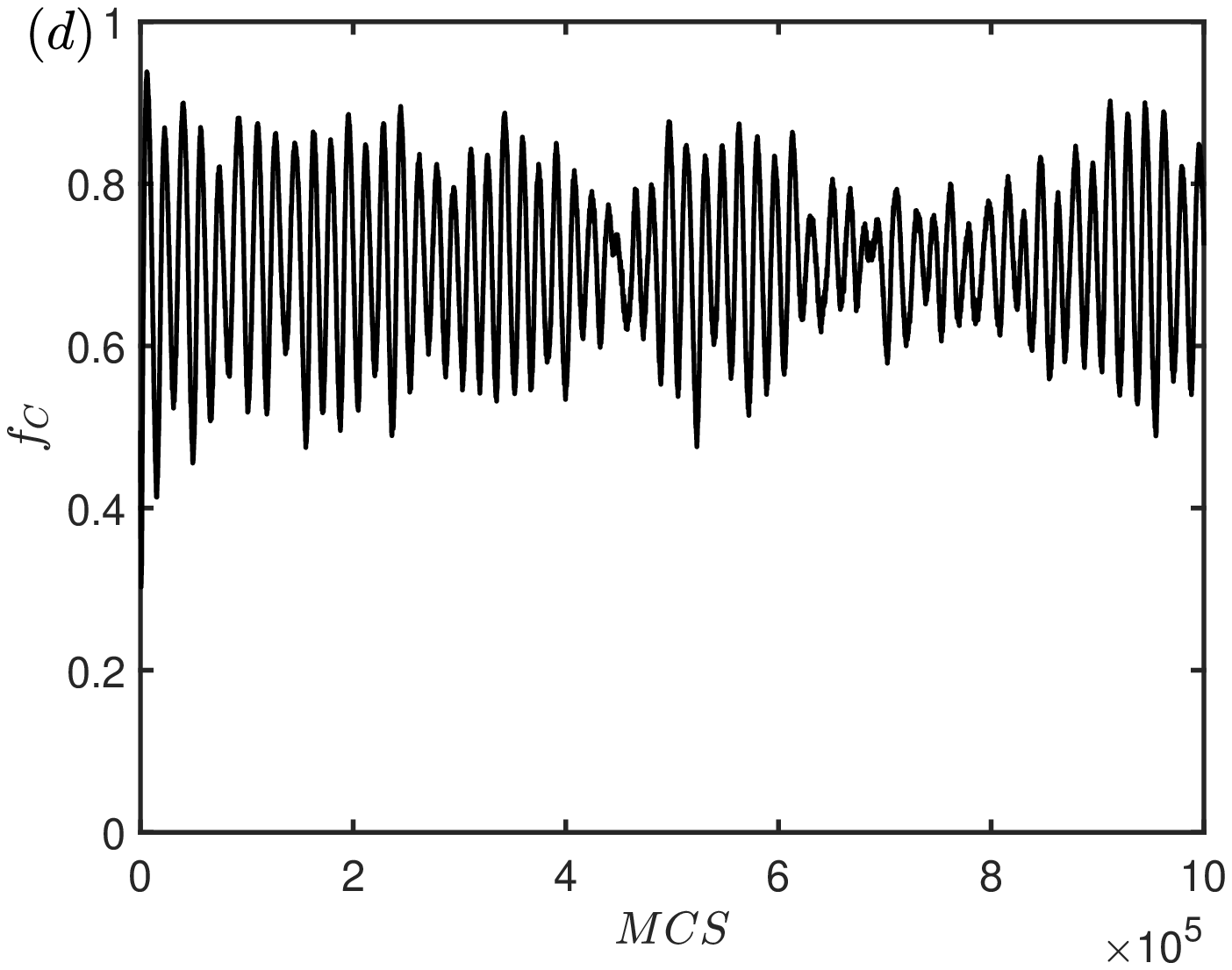}

\caption{
Typical time series of cooperation level $f_C$ for four different combination of the memory length $M$ and information acquisition probability $p$: (a) $M=1$, $p=1.0$; (b) $M=100$, $p=1.0$; (c) $M=20$, $p=0.2$; and (d) $M=100$, $p=0.2$. Parameter: $b=1.2$. 
}
\label{Fig:ts}
\end{figure*}

\section{Model}
\label{sec:model}
We study the prisoner's dilemma (PD) game with $N$ individuals that are located on an $L \times L$ square lattice with a periodic boundary condition. PD is a typical pairwise game for many social dilemmas, with the strategy being either cooperation (C) or defection (D). Mutual cooperation brings the reward $R$, mutual defection yields the punishment $P$ for each, and mixed encounter gives the cooperator the sucker’s payoff $S$, yet the temptation $T$ for the defector. The conditions of $T > R > P > S$ and $2R > T + S$ are required for PD. It's easy to see that mutual defection is the Nash equilibrium, even though the mutual cooperation is optimal for their collective profits. We adopt the weak version of PD~\cite{nowak1992evolutionary} in this work, $R = 1$, $P = S = 0$, and $T = b$, where $1.0 \le b \le 2.0$. 

To incorporate the history, each player (e.g. player $i$) has $k_i$ memories $\mathcal{M}_i(j)=\{\Pi_i^1(j),...,\Pi_i^M(j)\}$ of length $M$ to record the payoff information for each neighbor $j\in\Omega_i$, where $\Omega_i$ denote the neighborhood of player $i$, and $k_i=|\Omega_i|$ the degree of node $i$ and is four in the 2d square lattice. Due to the uncertainties in realistic surroundings, this information is not always accessible in each round. We assume that the payoff information in the neighborhood is obtained with an \emph{acquisition probability} $p$. The larger the value of $p$, the more likely to obtain their neighbors' payoff information, and vice versa. The case of $p=1$ recovers to the complete information scenario that many previous work assumed~\cite{szabo2007evolutionary,liu2010memory}.

Our model follows Monte Carlo (MC) simulation procedure. At the very start, each player's strategy is randomly assigned with either $C$ or $D$. The implementation of an elementary MC step is as follows. 
Firstly, we randomly select a player $i$ and one of its neighbor player $j$, and calculate their total payoffs $\Pi_{i,j}$ by playing with all their neighbors. 
Secondly, the obtained payoff of player $j$ is then recorded in the memory of player $i$ with the probability $p$;  if succeed $\Pi_{i}^1(j)=\Pi_{j}$, while the rest are shifted by one site with the earliest payoff information $\Pi_{i}^M(j)$ being removed accordingly; otherwise the memory $\mathcal{M}_i(j)$ will not be updated. 

Next,  the probability for player $i$ to adjust its strategy $s_i$ is given by the Fermi rule~\cite{szabo1998evolutionary}

\begin{equation}\label{eq:imitation}
\begin{aligned}
W(\rho_j^C\rightarrow s_i)=\dfrac{1}{1+\exp[(\overline{\Pi}_{i}-\overline{\Pi}_{j}')/K]},
\end{aligned}
\end{equation}

where $\overline{\Pi}_{i}$ is the averaged payoffs for player $i$ in the latest $M$ rounds since player $i$ knows exactly her/his own payoffs, and $\overline{\Pi}_{j}'=\sum_{m=1}^M \Pi_{i}^m(j)/M$ is the average payoffs of $j$ according to the memory $\mathcal{M}_i(j)$. Their subtraction in Eq. (\ref{eq:imitation}) can then be interpreted as the perceived payoff difference by player $i$. 
$K$ qualifies the surrounding noise during the imitation process, which is usually interpreted as the bounded rationality. $K\to 0$ indicates that players are absolutely rational and the imitation is deterministic; $K\to \infty$, on the other hand, means that the decision-making is completely random and is not affected by the neighbors. Here $K=0.1$ throughout the work~\cite{szabo1998evolutionary}. 
Instead of directly copying $s_j$ for imitation, player $i$ adopts $j$'s cooperation propensity defined as ${\rho}_j^C=N^C_j/M$, where $N^C_j$ is the number of times acting as a cooperator for player $j$ in its latest $M$ rounds. Different from the payoff information, the neighbours' strategy $s_j$ is always accessible for players $i$ because the two interact directly.
Finally, player $i$ behaves as a cooperator with the probability ${\rho}_j^C$ once the imitation is successful, otherwise no strategy change is made.

Notice that the above setup follows a typical asynchronous updating procedure. A complete Monte Carlo step (MCS) consists of $N$ elementary steps, meaning that every player updates its state exactly once on average. We compute the cooperation prevalence $f_C=\frac{1}{N}\sum_{i=1}^Ns_i$ as our primary order parameter, measuring the overall preference in cooperation of the population. If not stated otherwise, $L=100$ throughout our study, though the impact of size will be also discussed in the latter part. 

\begin{figure}[tbp]
\centering
\includegraphics[width=1.0\linewidth]{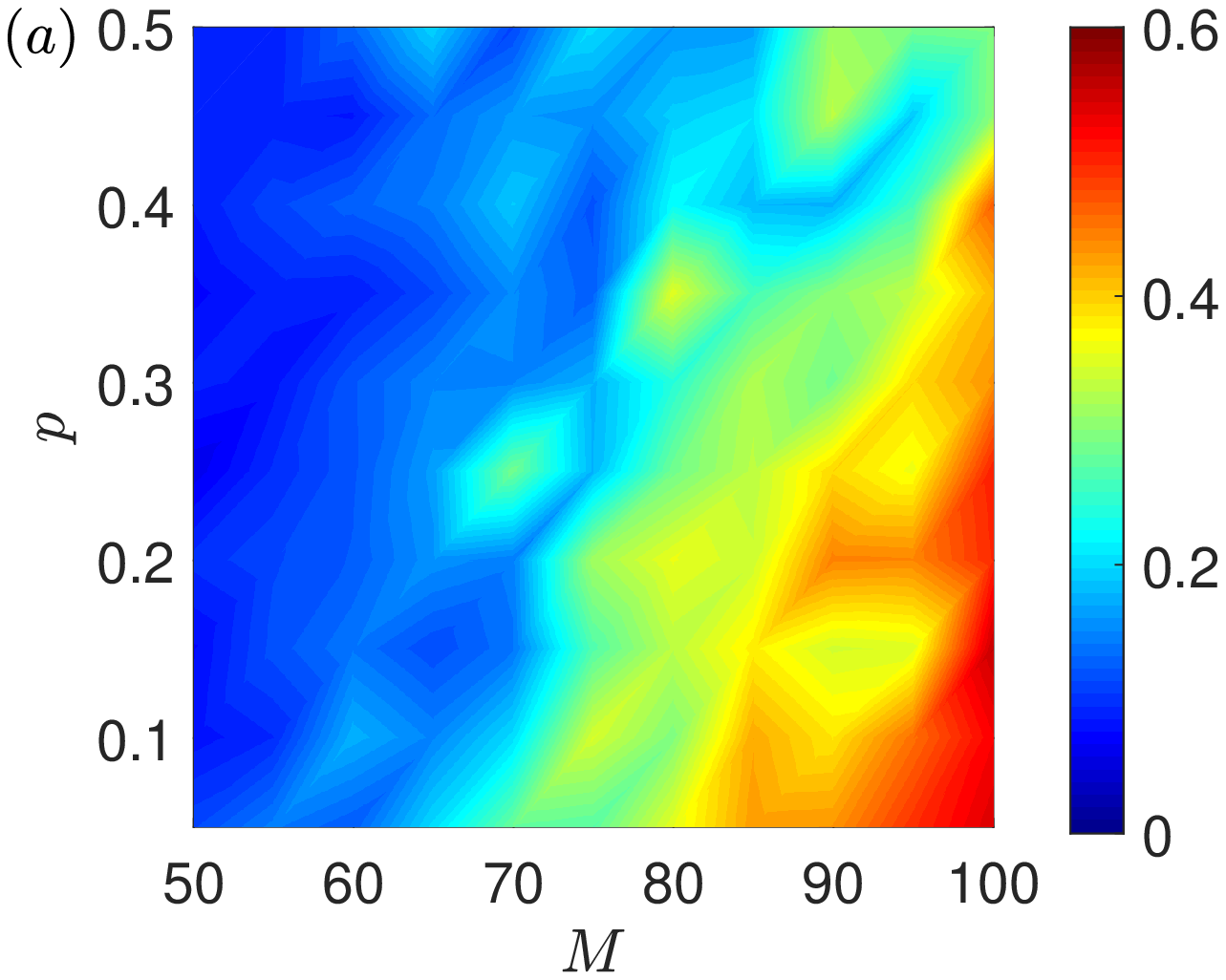}
\includegraphics[width=1.0\linewidth]{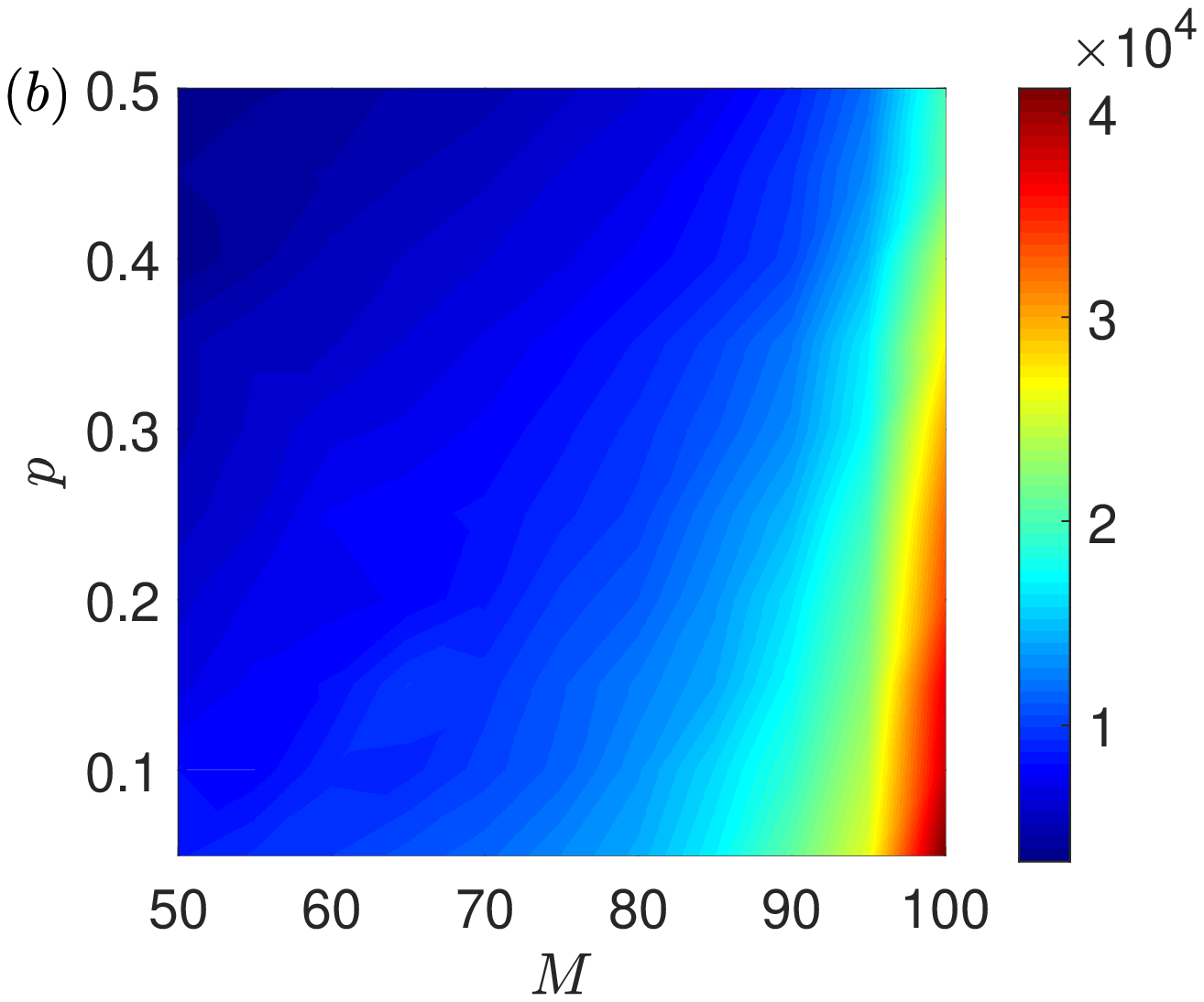}
\caption{ Color-coded phase diagram in the $M-p$ space.
(a) The difference between the average maximal and the minimal cooperation level $\langle f_C^{max}\rangle - \langle f_C^{min}\rangle$;
(b) The period corresponding to the principal frequency ($T=1/\omega_{max}$) by Fourier analysis of the time series for the given parameter combination of $M$ and $p$. 
Other parameter: $b=1.2$.
}
\label{Fig:phasediagram}
\end{figure}

\section{Numerical results}
\label{sec:Numerical results}

Figure~\ref{Fig:ts} reports typical time series for several combinations of information acquisition probability $p$ and memory length $M$ by fixing the temptation $b=1.2$. Let's first see the extreme case where the memory is absent ($M=1$ and $p=1$, Fig.~\ref{Fig:ts}(a)), which recovers the traditional Markov model~\cite{szabo2005phase}. As can be seen, no cooperation is seen for the given parameter $b$, in line with the previous result. When the non-Markov effect is incorporated but with full information acquisition ($M=100$ and $p=1$), cooperation arises and the fraction is above $80\%$, see Fig.~\ref{Fig:ts}(b). This means that by taking the history into account in the decision-making, the cooperation prevalence can be promoted, which is also reported in previous studies~\cite{liu2010memory}. But once the incomplete information is incorporated, the cooperation prevalence becomes unstable and its average value is slightly decreased, see Fig.~\ref{Fig:ts}(c,d). When the memory is short ($M=20$ and $p=0.2$, Fig.~\ref{Fig:ts}(c)), the cooperation prevalence $f_C$ shows noise-like fluctuations, while a longer memory ($M=100$ and $p=0.2$) gives rise to oscillatory behaviors, as shown in Fig.~\ref{Fig:ts}(d). A closer lookup shows this oscillation is quasi-periodic, confirmed by the pronounced peaks by doing Fourier analysis (see Sec. I in SM~\cite{SM}).\par
To further investigate the properties of the cooperation oscillation, we compute the difference between the average maximal prevalence at the peak $\langle f_C^{max}\rangle$ and the minimal prevalence at the valley $\langle f_C^{min}\rangle$ as a function of the memory length $M$ and the acquisition probability $p$. The obtained phase diagram Fig.~\ref{Fig:phasediagram}(a) confirms that oscillatory behaviors are more likely to occur for a longer memory and a small value of $p$, corresponding to strong non-Markov effect and severe incomplete information scenarios. Fig.~\ref{Fig:phasediagram}(b) plots the phase diagram of corresponding period $T=1/\omega_{max}$, where $\omega_{max}$ is the principal frequency in the power spectrum. It shows that the strongly non-Markov effect and severely incomplete information scenarios give rise to a long period of oscillation.\par
\begin{figure}[tbp]
\includegraphics[width=1.0\linewidth]{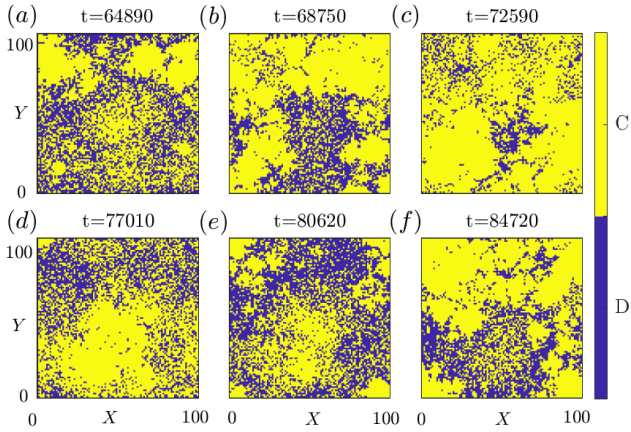}
\caption{
Cooperation state patterns at different times, where yellow and blue sites denote cooperators and defectors, respectively.  
From (a) to (c), cooperation is in the explosive stage, followed by the cooperation decline (d,e), and then rising again (f). 
Parameters: $M=100$, $p=0.2$, and $b=1.2$.
}
\label{Fig:pattern}
\end{figure}

To develop some intuition, typical spatial patterns are provided in Fig.~\ref{Fig:pattern} with the same parameters used as in Fig.~\ref{Fig:ts}(d). We can see that in Fig.~\ref{Fig:pattern}(a), where the cooperation level $f_c$ is at its minimum in time, some cooperator clusters emerge and start to grow. Afterwards, these clusters are merged with each other [Fig.~\ref{Fig:pattern}(b)], reaching its maximum at some time point [Fig.~\ref{Fig:pattern}(c)]. The later snapshots show that defectors increase but in a scattered manner [Fig.~\ref{Fig:pattern}(d)], and the cooperation prevalence $f_C$ declines to its minimum [Fig.~\ref{Fig:pattern}(e)]. Afterwards, the cooperator clusters re-emerge and grow [Fig.~\ref{Fig:pattern}(f)], the above process repeats again and again.\par
For completeness, we also provide the corresponding patterns for the cooperation propensity, the average cooperation level over the last $M$ rounds, see Fig.~\ref{Fig:pattern2}.
Compared to the scattered patterns shown in Fig.~\ref{Fig:pattern}, these plots show that the cooperation propensity varies smoothly across the domain, though the two figures are highly correlated. 
Detailed examination of the spatial patterns by compactness analysis shows that the cooperators are much closely clustered than the defectors, both of their compactness levels oscillate with the same frequency of $f_c$, but is not in perfect synchrony (see Sec. II in SM~\cite{SM}). 
 %indicates that the oscillation can be observed not only globally but also at the smaller scales, which is confirmed by monitoring $f_C$ at some randomly sampling locations .
%Provide the some local time series of f_C at smaller scale 
\section{Mechanism analysis}
\label{sec:Mechanism analysis}
To understand the formation of cooperation oscillation, we now turn to its dynamical mechanism analysis. Without loss of generality, we focus on the case with the parameters used in Fig.~\ref{Fig:ts}(d) in this section. 

\begin{figure}[tbp]
\includegraphics[width=1.0\linewidth]{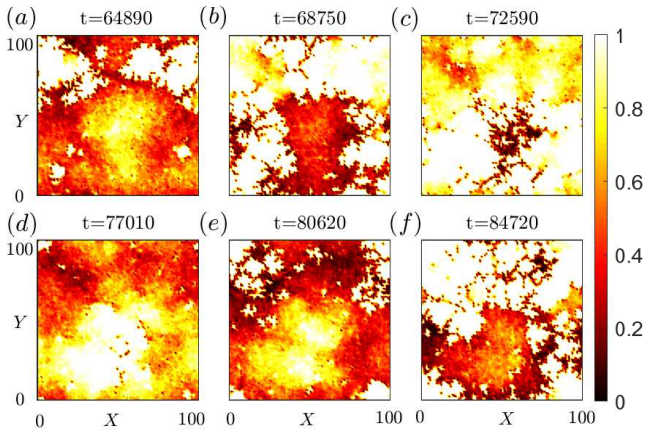}
\caption{
The corresponding cooperation propensity patterns defined as ${\rho}_j^C=N^C_j/M$. Other settings are exactly the same as Fig.~(\ref{Fig:pattern}). }
\label{Fig:pattern2}
\end{figure}

\begin{figure}[tbp]
\includegraphics[width=1.0\linewidth]{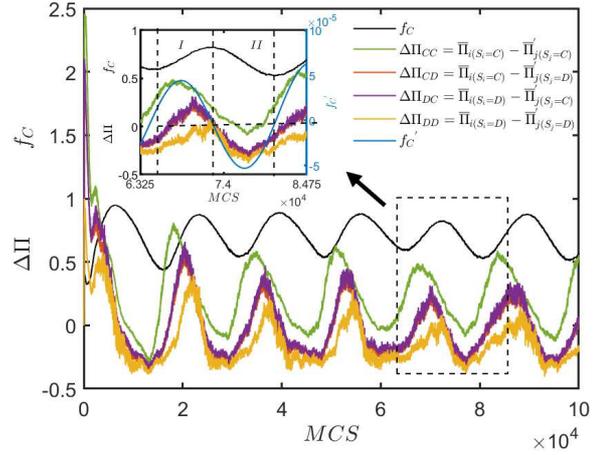}
\caption{
The time evolution of the average effective payoff difference for four different strategy combinations for $i-j$ pair, where $i$ is the focal node and $j$ is one of $i$'s neighbors. The inset shows an enlarged evolution section illustrating two phases: I -- cooperation explosion and II -- cooperation decline.
Parameters: $M=100$, $b=1.2$, and $p=0.2$.
}
\label{Fig:payoff_diff}
\end{figure}

Specifically, we monitor the effective payoff differences for different pairs to understand how they are correlated to the strategy update, and thus also the evolution of cooperation prevalence $f_C$, see Fig.~\ref{Fig:payoff_diff}.  For example, $\Delta\Pi_{CC}=\overline{\Pi}_i-\overline{\Pi}_j'$ denotes the difference between the average payoff of the focal player $i$ and the perceived payoff of its neighbor $j$, where both players are cooperators. By combination, there are four scenarios in total.
A critical observation in Fig.~\ref{Fig:payoff_diff} is that all four $\Delta\Pi$ are oscillating. According to the Eq.~(\ref{eq:imitation}), this implies that the probabilities to imitate are time-varying, their profiles are found to synchronously fluctuate with the derivatives of the cooperation prevalence $f'$ (see the inset in Fig.~\ref{Fig:payoff_diff}). Note that, different from the traditional case of directly imitating their neighbor strategy where C-C or D-D pairs remain unchanged by the mutual imitation, the updating in our case could still lead to the strategy change due to the probabilistic nature in the imitation. That's why we also need to monitor the cases of C-C and D-D pairs.

To see how the cooperation oscillates, we divide the evolution into two phases, see the inset in Fig.~\ref{Fig:payoff_diff}. 

\emph{Phase I --- cooperation explosion.}  This stage corresponds to Fig.~\ref{Fig:pattern}(a-c) and Fig.~\ref{Fig:pattern2}(a-c), where the cooperation prevalence is rising. The key feature is that the values of the four $\Delta\Pi$s are larger than their time averages, this is especially true for $\Delta\Pi_{CC}$. This means that the strategy updating in this stage is less frequent since the probability $W(\rho_j^C\rightarrow s_i)$ is relatively small. This is important for cooperation clusters to keep their high cooperation prevalence $\rho_C$. When cooperator clusters are formed, they are at an advantageous position over defectors, for the same logic behind the network reciprocity~\cite{nowak1992evolutionary}; therefore, the cooperation clusters expand and the low cooperation region shrinks as shown from Fig.~\ref{Fig:pattern}(a) to \ref{Fig:pattern}(c) and Fig.~\ref{Fig:pattern2}(a) to \ref{Fig:pattern2}(c).

\emph{Phase II --- cooperation decline}. Once the domain is dominated by the high cooperation prevalence [i.e. Fig.~\ref{Fig:pattern}(c) and Fig.~\ref{Fig:pattern2}(c)], shifting to defection would be a profitable move for individuals in such an environment. That's what happens from Fig.~\ref{Fig:pattern}(c)-\ref{Fig:pattern}(e) and Fig.~\ref{Fig:pattern2}(c)-\ref{Fig:pattern2}(e), where the cooperation prevalence declines. In this phase, all four $\Delta\Pi$ are lower than their time averages, meaning that the updating events occur very frequently.

At the end of phase II, the domain is now dominated by defectors, defectors are now at the disadvantage when they meet up with cooperation clusters. As a result, the cooperation decline stops, the regions with low $\rho_C$ are invaded again by cooperation clusters, and the dynamical process of Phase I restarts, and repeats again and again. 

In both phases, the presence of incomplete information (i.e. the acquisition probability $p<1$) facilitates the cooperation explosion and decline, respectively. 
In Phase I, a small $p$ inhibits the spread of defection within the cooperation clusters. Consider the scenario where a player (say $j$) defects within a highly cooperative neighborhood, this brings $j$ a high payoff as a result, but due to the presence memory and the acquisition probability, this payoff information is not immediately reflected in the payoff difference perceived by its neighbors. The payoff advantage of defection needs a longer time to spread within the cooperation clusters; this leaves a time window for cooperation cluster's growth till a relatively high level of cooperation.
In Phase II, our scenario facilitates the spread of defection on the contrary.  Consider the invasion of defection (say player $j$) into regions with a relatively high cooperation [say player $i$, c.f. Fig.~\ref{Fig:pattern2}(d)], the invasion is not possible in the absence of incomplete information because defectors have lower payoffs when they are clustered, showing no advantage in this scenario. But what player $j$ perceived at this time point is higher $\overline{\Pi}_j'>\overline{\Pi}_j$ because $\overline{\Pi}_j'$ includes early decent payoffs when $j$ was also a cooperator, this helps its invasion and leads to the extinction of relatively high cooperation regions.
In both phases, the degree of cooperation explosion and decline varies due to the stochastic nature, the amplitude of oscillation is fluctuating, as we seen in Fig.~\ref{Fig:ts}(d).

\begin{figure}[htbp]
\includegraphics[width=1.0\linewidth]{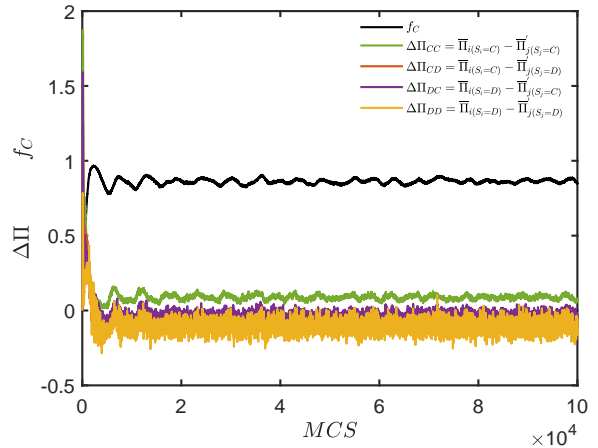}
\caption{
The time evolution of the average effective payoff difference for four different strategy combinations for $i-j$ pair, where $i$ is the focal node and $j$ is one of $i$'s neighbors. 
Parameters: $M=100$, $b=1.2$, and $p=1.0$. 
}
\label{Fig:without_II}
\end{figure}

In brief, due to the presence of incomplete information ($p<1$), there is a perception mismatch that inhibits the invasion of defection in the first phase and facilitates its invasion in the second phase. That leads to a cooperation explosion and followed by a cooperation decline, and the process repeats, oscillation is thus formed. The longer the memory, the stronger mismatch we expected. That explains the oscillatory behaviors are only observed in the case of a large $M$ and a small $p$.
As a comparison, when the incomplete information is removed from the model ($p=1$), the memory effect alone is insufficient to trigger the cooperation oscillation, the payoff difference is relatively stable without perception mismatch, see Fig.~\ref{Fig:without_II}. Their spatial patterns show no cooperation
explosion and decline after transient evolution, see see Sec. III in SM~\cite{SM}.

\section{Discussion and Conclusion}
\label{sec:conclusion}

In summary, our work extends the previous game-theoretic model of cooperation, by allowing for both the memory effect and the incomplete information. While the former factor is well-motivated and has been extensively studied previously~\cite{wang2006memory,qin2008effect,yong2010payoff,jiang2009reducing,szabo2007evolutionary}, the latter considers the fact that players in the real world are not always aware of their neighbors' payoff information, not the complete information scenario assumed as in most previous work~\cite{szabo2007evolutionary}. The inclusion of memory leads to the promotion of cooperation prevalence as also revealed previously, we uncover here that further inclusion of the incomplete information de-stabilizes the lifted prevalence and yields a quasi-periodic oscillation. The oscillation is strengthened for a long memory and a strong incomplete information situation.
Further analysis shows that the oscillation is an intrinsic property when both ingredients are included. They together lead to a misperception of individual neighborhood's payoff, erroneously making decisions. The consequence is that the invasion of defection is inhibited and facilitated in the phase of cooperation explosion and decline, respectively, causing cooperation oscillation as a result. 

Since the oscillation is an intrinsic property of the system, and no synchrony is found in the prevalence for different locations, we expect that the oscillation amplitude of the cooperation prevalence $f_C$ becomes smaller as the system size becomes larger. This is confirmed in our further experiments (see Sec. IV in SM~\cite{SM}). From the perspective of statistical physics, the observation of oscillation in $f_C$ is a finite-size effect, and cannot be expected in an infinitely large system. However, if we focus on a local region of the population, oscillation is still observable. Since the populations in the real world are always of finite size, the oscillatory behaviors should still be observed. When the population is structured in complex networks, numerical experiments show qualitatively the same phenomena (data not shown). However, once the game is replaced with the snowdrift game, no oscillation is seen, this is because the failure of spatial reciprocity in this game, where cooperators are not well-clustered for their changes~\cite{hauert2004spatial}.

Compared to the existing game-theoretic models for explaining oscillatory behaviors, our pairwise game model does not require three or more species~\cite{szolnoki2014cyclic}, or seek for other complex mechanisms such as mobility or conformity~\cite{reichenbach2007mobility,yang2022oscillation}. In this sense, our model provides a relatively simple framework to understand the emergence of oscillation. Given the omnipresence of incomplete information scenario and the non-Markov process in the realistic games, our model may provide a plausible explanation for a range of oscillatory phenomenon. 

\begin{acknowledgements}
We are supported by the Natural Science Foundation of China under Grants Nos. 12075144 and 61703257. ZJQ is supported by the Natural Science Foundation of China under Grants No. 12165014.
\end{acknowledgements}

\paragraph{Data availability}
The datasets generated during the current study are available from the corresponding author on reasonable request.

\section*{Declarations}

\paragraph{Conflict of interest}
The authors declare that they have no conflict of interest.

\bibliography{ref}   % name your BibTeX data base

\end{document}